# Dual-Gradient Plasmonic qBIC Metasurface for Time-Resolved *In Situ* Optimization of Molecular Vibrational Sensing in Water

Tao Jiang[1,2], Michael Hirler[1,2], Lina Rohrer[1,2], Martin Barkey[2], Dmytro Gryb[2], Leonardo de S. Menezes[2,3], Silvia Holler[4], Stefan A. Maier[5,6], Yohan Lee*[1,2], Alexander A. Antonov†[1,2], Andreas Tittl‡[1,2]

[1]Institute of Photonics, Hamburg University of Technology, 21073 Hamburg, Germany

[2]Chair in Hybrid Nanosystems, Nanoinstitute Munich, Faculty of Physics, Ludwig-Maximilians-University, 80539 München, Germany

[3]Departamento de Física, Universidade Federal de Pernambuco, 50670-901 Recife-PE, Brazil

[4]Cellular Computational and Integrative Biology Department, University of Trento, 38123 Trento, Italy

[5]School of Physics and Astronomy, Monash University, Clayton, 3800 Victoria, Australia

[6]Department of Physics, Imperial College London, SW7 2BW London, UK

Email: * yohan.lee@tuhh.de / † alexander.antonov@tuhh.de / ‡ andreas.tittl@tuhh.de

## ABSTRACT

Metaphotonic platforms based on quasi-bound states in the continuum (qBICs), harnessing strong near-field enhancement and deeply subwavelength field confinement, provide a versatile framework for surface-enhanced infrared absorption (SEIRA) spectroscopy. However, real-time molecular sensing in water using qBIC metasurfaces remains challenging. First, strong water absorption in the mid-infrared region damps qBICs and obscures weak analyte signals. Second, conventional metasurfaces rely on discrete arrays targeting individual wavelengths and coupling conditions, increasing the device footprint and fabrication effort required to match the qBIC to the molecular vibrational resonance. Here, we present a dual-gradient plasmonic qBIC metasurface with diamond-shaped resonators for *in situ* molecular vibrational sensing in water, where spatially encoded gradients in the scaling factor and asymmetry parameter independently control the qBIC spectral positions and radiative rates, respectively, across a 1000 μm × 700 μm footprint. Building on this capability, we experimentally monitor lipid vesicle dynamics in real time and resolve the carbonyl vibrational signature despite the unavoidable water absorption. We then directly identify the optimal sensing condition and track its evolution across the spatially encoded parameter space. The results establish a compact single-chip strategy that combines adsorption-kinetics monitoring, vibrational fingerprint detection, and on-chip optimization, opening opportunities for investigating biological dynamics under aqueous conditions.

## INTRODUCTION

Mid-infrared (mid-IR) spectroscopy provides label-free and nondestructive access to molecular detection and identification by probing inherent molecular vibrations that encode chemical composition and conformation.[1,2] This capability has made mid-IR spectroscopy an important tool in drug development and clinical biomedical diagnostics.[3] Because molecular vibrational absorption scales with analyte thickness in accordance with the Beer–Lambert law,[4] conventional mid-IR spectroscopy often cannot detect molecules at very low concentrations or in nanometer-scale layers.[5,6] Surface-enhanced infrared absorption (SEIRA) spectroscopy overcomes this by exploiting strongly localized electromagnetic fields at subwavelength scales,[7,8] substantially enhancing the mid-IR absorption of molecules and achieving high detection sensitivity.[9–11]

Resonant metasurfaces have become a central architecture for SEIRA sensors.[12] The optical response of resonant metasurfaces can be tailored through material selection, spanning dielectric,[13,14] plasmonic[15,16] or hybrid structures,[17,18] as well as resonator geometry.[19,20] This versatility has enabled high-performance biosensing and advanced functionalities[21], including simultaneous quantification and identification of molecules.[17,22,23] In most reported SEIRA implementations, analytes are measured in the dry state, either immobilized as molecular monolayers after incubation and rinsing,[24,25] deposited by spin-coating,[26,27] or concentrated through drop-casting and evaporation-assisted processes.[28]

Implementing SEIRA with resonant metasurfaces directly in aqueous environments is highly desirable, since it gives access to biomolecular processes under physiologically relevant conditions. However, it remains challenging because the weak absorption of biomolecules is often obscured by intense water absorption bands in the mid-IR, particularly when the biomolecular signatures overlap spectrally with these bands. Previous studies have addressed this by measuring metasurfaces under a thin residual water film[29] or replacing water with $D_2O$,[30] whose bending vibration mode shifts away from the amide I-II region. Both workarounds have drawbacks: residual water films are sensitive to environmental factors such as humidity and temperature,[31] while $D_2O$ does not fully reproduce native hydration conditions and may perturb hydrogen-bonding networks, molecular dynamics, and the stability of biomolecules.[32] To advance biosensing in water, mid-IR plasmonic metasurfaces with gold nanorods have been explored, which benefit from the lightning-rod effect associated with sharp geometric features,[33,34] thereby enhancing the absorption signals of immobilized biomolecules in aqueous media.[35–37] In these configurations, illumination from the substrate side minimizes the optical path length through water and thus avoids complete attenuation of the infrared signal.

Photonic resonances, specifically originating from symmetry-protected bound states in the continuum (BICs),[38,39] offer distinct advantages for improving mid-IR molecular sensing.[13,40,41] Through controlled symmetry breaking, BICs can be tuned into radiatively accessible quasi-BICs (qBICs), with linewidths and far-field coupling strengths determined by the degree of structural asymmetry,[42–45] thereby allowing the coupling between the photonic resonance and molecular vibrational modes to be tailored.[20,46] Because molecular vibrational fingerprints are discrete and chemically specific, matching the spectral position and optimizing the interaction strength between qBICs and the target resonance has typically required multiple separate arrays, [27] multi-resonant antennas,[23] iterative design optimization,[47] tunable resonances,[37,48] or angle-dependent measurements[49] to identify the optimal sensing condition.

The recently introduced concept of gradient dielectric metasurfaces, in which qBIC spectral positions and radiative lifetimes vary spatially, offers an efficient strategy for single-chip sensing.[25,50,51] In liquid environments, however, the mode volume of dielectric qBICs extends substantially into the surrounding medium, which increases susceptibility to water-induced damping and limits sensing performance. Plasmonic qBICs provide a complementary route through stronger surface-localized fields, which substantially decrease the mode volume, and are therefore better suited for mid-IR molecular sensing in water, as supported by our recent comparative study of plasmonic and dielectric qBICs.[52] Nevertheless, *in situ* molecular vibrational sensing supported by qBICs in water and the real-time identification of the optimal sensing condition remain unexplored.

Here, we demonstrate a dual-gradient plasmonic qBIC metasurface for real-time *in situ* optimization of molecular vibrational sensing in water on a single chip. First, by incorporating gradients in both the scaling factor and asymmetry parameter, we spatially encode qBIC spectral positions and radiative decay rates, respectively, enabling broad spectral coverage and systematic optimization of the coupling strength. We replace the elliptical resonators with a diamond-shaped geometry, which further strengthens the near-field enhancement through the lightning-rod effect at the sharp corners. We then experimentally demonstrate *in situ* vibrational sensing of lipid vesicles in water. Notably, the dual-gradient plasmonic metasurface identifies the resonator configuration that maximizes the carbonyl vibrational response, even under strong aqueous background absorption. These results establish the dual-gradient plasmonic metasurface as a single-chip approach that integrates monitoring of molecular adsorption kinetics, vibrational fingerprint characterization, and on-chip optimization of sensing performance, thereby providing a compact route to investigating real-time biological processes in aqueous environments.

## RESULTS AND DISCUSSION

### Concept of a Dual-Gradient Plasmonic qBIC Metasurface

The mid-IR sensing chip is a dual-gradient gold metasurface fabricated on a $CaF_2$ substrate and operated in water, where a thin analyte layer adsorbs onto resonators during sensing (Figure 1a). Each unit cell comprises a pair of diamond-shaped plasmonic nanorods with lateral dimensions $S \cdot d_1$ and $S \cdot d_2$, arranged with lattice periods $S \cdot P_x$ and $S \cdot P_y$, where $S$ is the in-plane scaling factor. The gold and analyte layer thicknesses are denoted by $h_{Au}$ and $h_{an}$, respectively (Figure 1b). The incident light is polarized along the $y$ direction. Structural asymmetry is introduced by rotating the two diamond nanorods in opposite directions by an opening angle $\theta$. The asymmetry parameter $\theta$ controls the radiative decay rate of the qBIC, which determines its linewidth and coupling strength to the far field. Compared with conventional elliptical nanorods,[13,50] the diamond-shaped resonators exhibit a 30.8% increase in $Q$-factor from 11.7 to 15.3 at $\theta = 12.5°$, an average increase of 32.2% over the asymmetry range of $\theta = 7.5°–35°$, and stronger near-field enhancement with localized hot spots at the resonator tips (Figure 1c and Supplementary Note 1). As the SEIRA response is governed by the local electric-field intensity,[7] this enhanced near field strengthens the interaction with molecular vibrations and facilitates the detection of weak absorption signatures in aqueous environments. Accordingly, the diamond geometry yields an 8.9% relative increase in absorbance at $\theta = 12.5°$ and an average increase of 12.5% across the same asymmetry range (Supplementary Note 1).

To simultaneously match the spectral position and radiative decay rate of the qBIC to the molecular vibrational resonance, we introduce a dual-gradient metasurface in which the scaling factor $S$ and asymmetry parameter $\theta$ are varied along two orthogonal directions (Figure 1d). The scaling factor $S$ sets the resonance position and continuously tunes the qBIC across the molecular vibrational band, whereas the asymmetry parameter $\theta$ adjusts the radiative loss of the qBIC and controls its interaction strength with the molecular vibrational mode.[50] Representative scanning electron microscopy (SEM) images and the corresponding changes in the reflectance signal upon varying the scaling factor and asymmetry parameter are shown in Figure 1e. Thus, a single compact metasurface provides a full set of resonator configurations, rather than requiring multiple separately fabricated arrays.[30]

The molecular vibrational response is obtained by comparing the reflectance spectra before and after analyte adsorption on the gold metasurface in water (Figure 1f). The absorbance is calculated as $A = -\log(R_w/R_{wo})$, where $R_w$ and $R_{wo}$ denote the reflectance with and without analyte, respectively. During analyte adsorption, analyte molecules progressively replace water within the metasurface near-field region, modifying the local refractive index and shifting the qBIC spectral position. This produces a smooth broadband absorbance background, while the molecular vibrational absorption of the analyte appears as localized spectral features superimposed on it (Supplementary Note 2). We define the baseline-corrected absorbance as $A_{bc} = A - A_b$, where $A_b$ is the background component. Subtracting $A_b$ removes the refractive-index-induced change and isolates the molecular vibrational contribution. The integrated absorbance $\beta$ is obtained by integrating $A_{bc}$ over the vibrational band between wavenumbers $\nu_1$ and $\nu_2$. During the *in situ* sensing experiment, the dual-gradient gold metasurface maps the real-time evolution of analyte adsorption across the two-dimensional parameter space defined by the scaling factor and asymmetry parameter (Figure 1g). As the molecular coverage increases with time, the integrated absorbance evolves from a weak initial response into a distinct maximum at the best-matched resonance position and coupling conditions, providing a direct route to track molecular sensing dynamics under the optimal configuration.

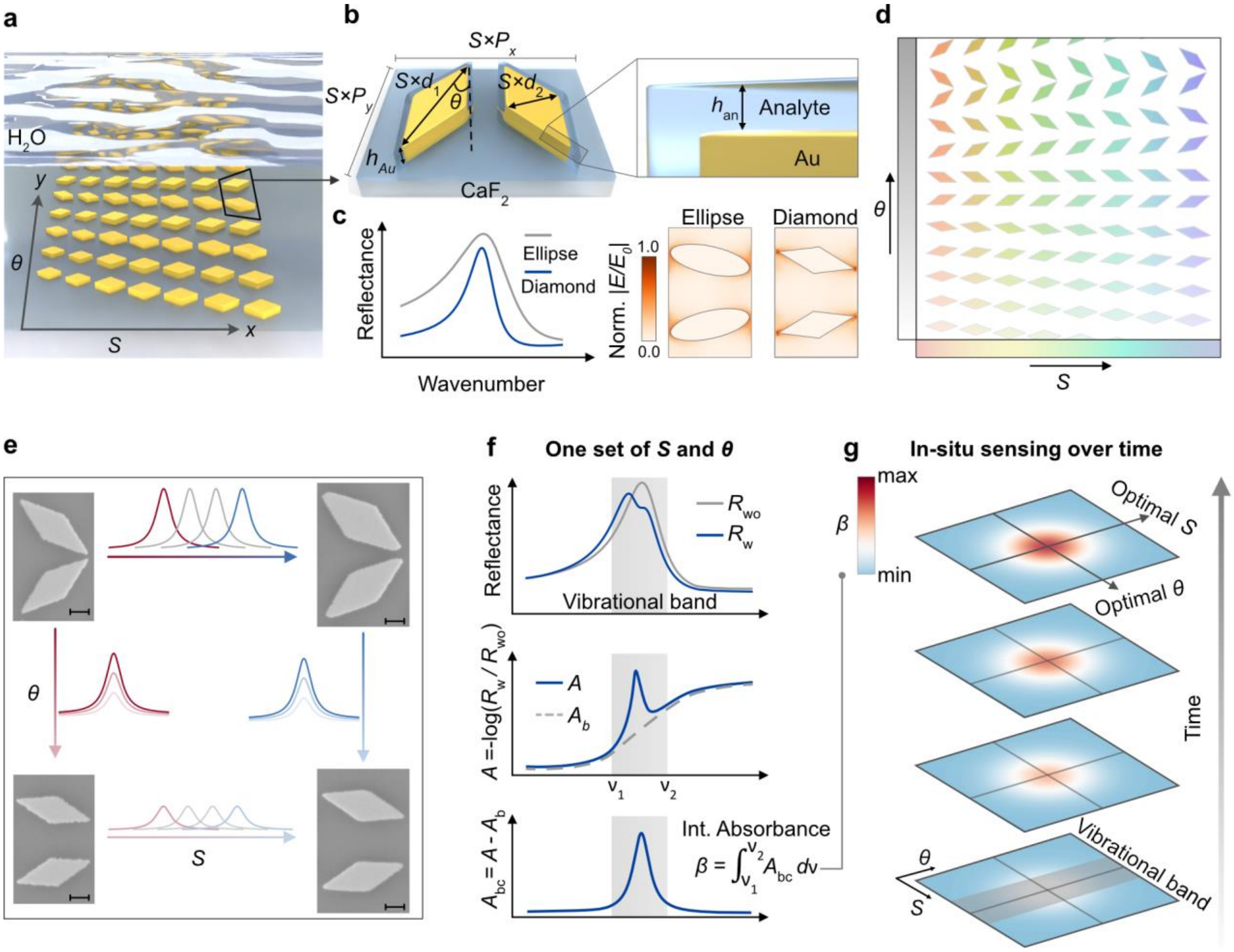

**Figure 1.** Concept of a single-chip dual-gradient plasmonic qBIC metasurface for *in situ* molecular sensing in water. (a) Schematic of the dual-gradient plasmonic qBIC metasurface covered by a thin analyte layer and immersed in water. (b) Unit cell geometry of the metasurface. (c) Replacing conventional elliptical nanorods with diamond-shaped resonators yields a higher quality factor (*Q*-factor) and stronger near-field enhancement at the resonator tips. (d) Schematic of the metasurface, incorporating a spectral gradient defined by the scaling factor $S$ along the horizontal direction ($x$-direction) and a coupling-strength gradient defined by the asymmetry parameter $\theta$ along the orthogonal direction ($y$-direction). (e) Scanning electron microscopy (SEM) images of representative unit cells and the corresponding reflectance spectra. Scale bar: 500 nm. (f) Reflectance spectra of a representative set of parameters ($S$, $\theta$) without ($R_{wo}$) and with ($R_w$) the analyte. The absorbance is calculated as $A = -\log(R_w/R_{wo})$, and the baseline-corrected absorbance is defined as $A_{bc} = A - A_b$, where $A_b$ is the baseline contribution from the refractive-index change as water is exchanged for analyte. (g) Concept of *in situ* real-time sensing using the dual-gradient plasmonic metasurface. The integrated absorbance $\beta$ is obtained by integrating $A_{bc}$ over the analyte vibrational band between wavenumbers $v_1$ and $v_2$. As analyte adsorption progresses, the integrated absorbance increases over time, with the strongest response localized in the central region

corresponding to the optimal $S$ and $\theta$. The shaded region in the lower panel marks the molecular vibrational band.

**Numerical Modeling of a Dual-Gradient Plasmonic qBIC Metasurface**

To model sensing performance, both water and the analyte were treated as homogeneous, dispersionless lossy media. Water was assigned a refractive index of $n$ = 1.34 and an extinction coefficient of $k$ = 0.03,[53] while the analyte was modeled with $n_{an}$ = 1.45 and $k_{an}$ = 0.25 as a 5 nm thick conformal layer on the gold metasurface immersed in water, comparable to the thickness of a supported lipid bilayer.[54–57] To analyze real-time sensing performance, we introduce an analyte coverage ratio ($CR$) based on an effective-medium approximation.[58] The analyte $CR$ is defined as the fraction of the near-field sensing region occupied by analyte molecules, ranging from 0 to 1, where the sensing region is represented by the conformal analyte layer. $CR$ = 0 corresponds to a fully water-filled near-field region, whereas $CR$ = 1 represents complete analyte coverage. An increasing analyte $CR$ describes the progressive replacement of water by analyte and thus represents the evolution of adsorption during *in situ* sensing. The effective permittivity of the sensing region is complex and given by $\varepsilon_{\mathrm{eff}} = CR\cdot\varepsilon_{\mathrm{an}} + (1 - CR)\cdot\varepsilon_{\mathrm{H2O}}$, where $\varepsilon_{\mathrm{an}} = \varepsilon_{1,\mathrm{an}} + i\cdot\varepsilon_{2,\mathrm{an}}$ and $\varepsilon_{\mathrm{H2O}} = \varepsilon_{1,\mathrm{H2O}} + i\cdot\varepsilon_{2,\mathrm{H2O}}$ are the complex permittivities of the analyte and water, respectively.

In this simplified model without material dispersion, variations in the scaling factor $S$ have a negligible effect on the absorbance, since the scaling factor primarily shifts the spectral position of the qBIC. For a given lossy environment, the absorbance response depends predominantly on the radiative loss, which is controlled by the asymmetry parameter $\theta$.[52] The influence of $\theta$ on the plasmonic qBIC was examined by calculating the reflectance spectra for different asymmetry parameters under two conditions: without analyte ($CR$ = 0) and with a partial analyte coverage ratio ($CR$ = 0.8), with the latter selected as a representative intermediate coverage that produces a clear yet unsaturated modulation of the reflectance (Figure 2a,b). As $\theta$ increases from 10° to 40°, the reflectance amplitude $R_{\mathrm{wo}}$ increases substantially, indicating enhanced radiative coupling of qBICs to the far field. When the analyte $CR$ increases from 0 to 0.8, the qBIC redshifts and the reflectance amplitude decreases. This occurs because analyte molecules replace water molecules near the gold resonators, introducing both a higher local refractive index and larger intrinsic loss to the qBIC. We extracted the maximum reflectance for each asymmetry parameter under both $CR$ = 0 and $CR$ = 0.8 (Figure 2b), showing that the reflectance difference between the two conditions grows with increasing $\theta$.

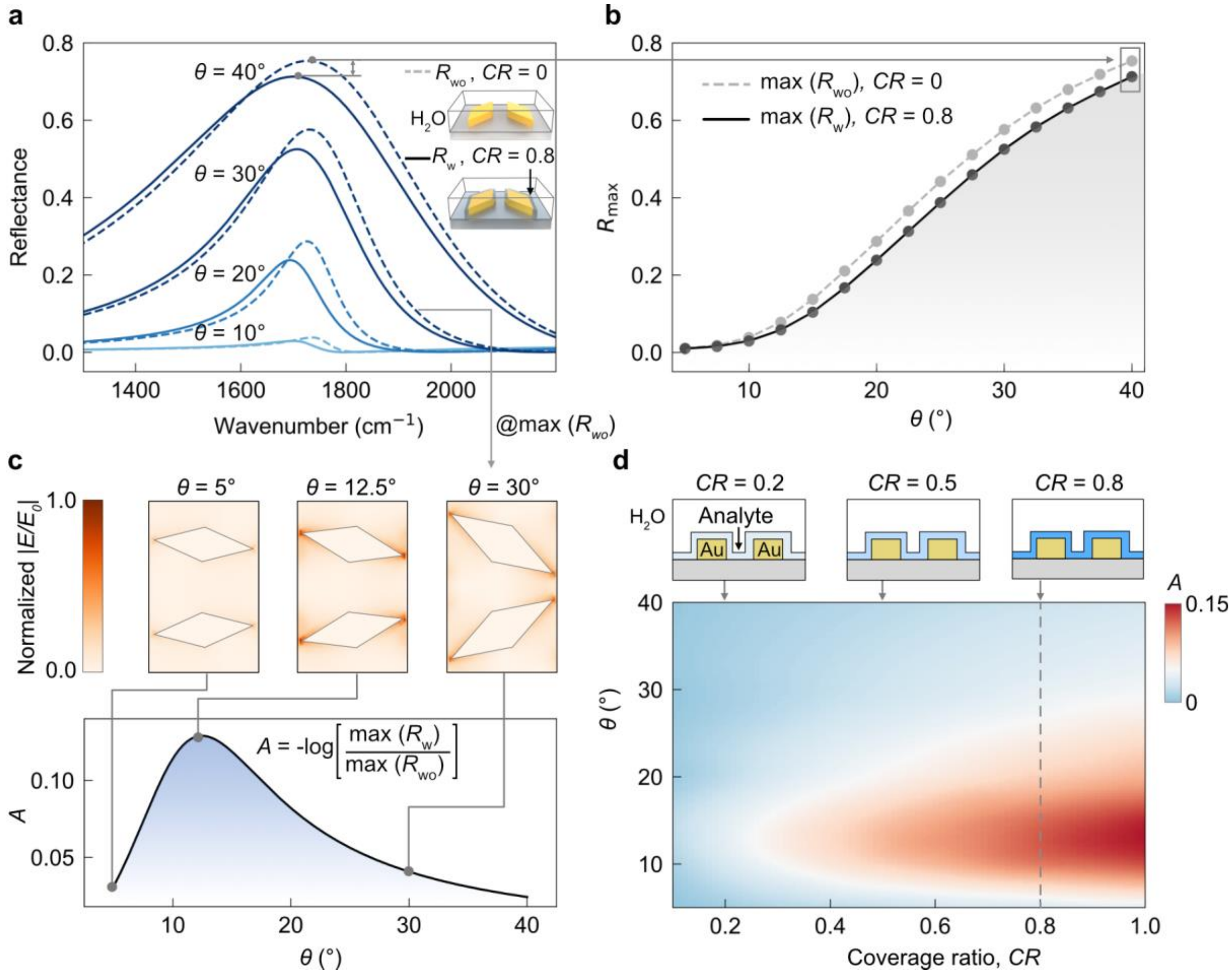

**Figure 2.** Numerical studies of metasurface sensing performance in water. The analyte coverage ratio (*CR*) is introduced to quantify the fraction of the metasurface near-field region occupied by the analyte in the liquid environment. (a) Reflectance spectra of the metasurface with different opening angles ($\theta$ = 10°, 20°, 30°, and 40°) under two conditions: without analyte (*CR* = 0) and with partial analyte coverage (*CR* = 0.8). Insets show the corresponding unit cells immersed in water, indicated by the surrounding box. (b) Maximum reflectance as a function of the asymmetry parameter $\theta$ for the coverage ratios *CR* = 0 and *CR* = 0.8. (c) Absorbance *A* and the corresponding normalized near-field enhancement as a function of the asymmetry parameter $\theta$. (d) Absorbance map as a function of the coverage ratio *CR* and asymmetry parameter $\theta$. Insets illustrate the metasurface at different analyte *CR*s in water.

Figure 2c presents the absorbance *A* together with the corresponding near-field enhancement maps, which show hot spots at the resonator tips for different asymmetry parameters $\theta$. The absorbance reaches a maximum at $\theta$ = 12.5°, which reflects a trade-off between radiative coupling of qBICs and intrinsic losses in the system. At an asymmetry parameter of 5°, the qBIC remains weakly coupled to the far field and is damped by intrinsic losses from gold and the surrounding environment, resulting in a low signal-to-noise ratio. Conversely, increasing the asymmetry parameter ($\theta$ = 30°) introduces excessive radiative loss, which broadens the resonance and weakens the near-field enhancement.

The dependence of absorbance on both the asymmetry parameter $\theta$ and analyte *CR* was then mapped to identify the optimal sensing condition during analyte adsorption (Figure 2d). As the analyte *CR* increases, the absorbance grows because a larger fraction of the water is replaced by analyte in the qBIC near-field region, leading to stronger signal modulation. The maximum absorbance remains centered around $\theta = 12.5°$ across a broad range of analyte *CR*s, indicating that the optimal asymmetry parameter is nearly independent of analyte coverage. This highlights the critical role of qBIC radiative losses in optimizing molecular vibrational sensing in a given lossy environment.

## Experimental Realization of a Dual-Gradient Plasmonic qBIC Metasurface

To verify the control of the plasmonic qBIC by the asymmetry parameter $\theta$ and scaling factor $S$, we fabricated the dual-gradient gold metasurface with a footprint of 1000 µm × 700 µm on a $CaF_2$ substrate (Methods section) and characterized the metasurface in water without analyte. Varying $\theta$ changes not only the radiative losses of the qBIC but also its spectral position. The diamond nanorod geometry was therefore corrected, following an established design methodology, to keep the resonance position fixed,[50] where only the dimensions $d_1$ and $d_2$ were adjusted (Supplementary Note 3). Hyperspectral reflectance imaging was performed using a mid-IR microscope (Spero, Daylight Solutions), with the reflectance response recorded at each spatial position of the dual-gradient plasmonic metasurface as a function of wavenumber. All spectra were collected from the substrate side of the metasurface to minimize water absorption along the optical path. Representative reflectance snapshots from 1550 to 1800 $cm^{-1}$ with a step size of 50 $cm^{-1}$ reveal a clear spatial distribution of the resonant response across the metasurface (Figure 3a for reflectance in water and Supplementary Note 4 for reflectance in air). Supplementary Videos 1 and 2 present the reflectance response of the metasurface during the spectral scan in air and water, respectively, with each frame corresponding to a single wavenumber. A local suppression of reflectance is evident near 1650 $cm^{-1}$, resulting from the overlapping water bending mode.

To examine the role of the asymmetry parameter, reflectance spectra were extracted along the $\theta$-gradient direction ($y$-direction) while keeping the scaling factor fixed at $S = 0.87$ (Figure 3b for reflectance in water and Supplementary Note 5 for reflectance in air). Additional results at $S$ = 0.84 and 0.90 are shown in Figure S6. As the asymmetry parameter $\theta$ increases from 10° to 36°, the reflectance amplitude increases substantially, while the spectral position remains nearly unchanged. Meanwhile, the water-absorption-induced dip near 1650 $cm^{-1}$ becomes more prominent due to the enhanced interaction between the qBICs and the aqueous environment. The corresponding reflectance map further confirms that the qBICs are well aligned spectrally along the $\theta$-gradient direction ($y$-direction) of the metasurface array, while their reflectance amplitude is systematically controlled by the asymmetry parameter $\theta$ (Figure 3c).

The effect of the scaling factor was evaluated by extracting spectra along the $S$-gradient direction ($x$-direction) at a fixed asymmetry parameter of $\theta = 20°$ (Figure 3d for reflectance in water and Supplementary Note 7 for reflectance in air). Results for fixed asymmetry parameters $\theta$ of 10°, 30°, and 40° are shown in Figure S8. At $\theta = 10°$, the qBICs are strongly suppressed in water and exhibit only a weak reflectance response, whereas at $\theta = 40°$ they retain high reflectance amplitudes, decreasing by approximately 0.1 compared with those in air. Increasing the scaling factor $S$ from 0.75 to 0.95 leads to a redshift of the qBIC across the measured spectral window.

The corresponding reflectance map shows a continuous spectral coverage of the qBIC as a function of the scaling factor *S*, except for the interruption near 1650 cm$^{-1}$ caused by water absorption (Figure 3e). This spectral gradient facilitates alignment of the qBIC with the target molecular vibrational region. Thus, our measurements experimentally validate the dual-gradient design in water, where the metasurface continuously maps both coupling strength and spectral position on a single chip.

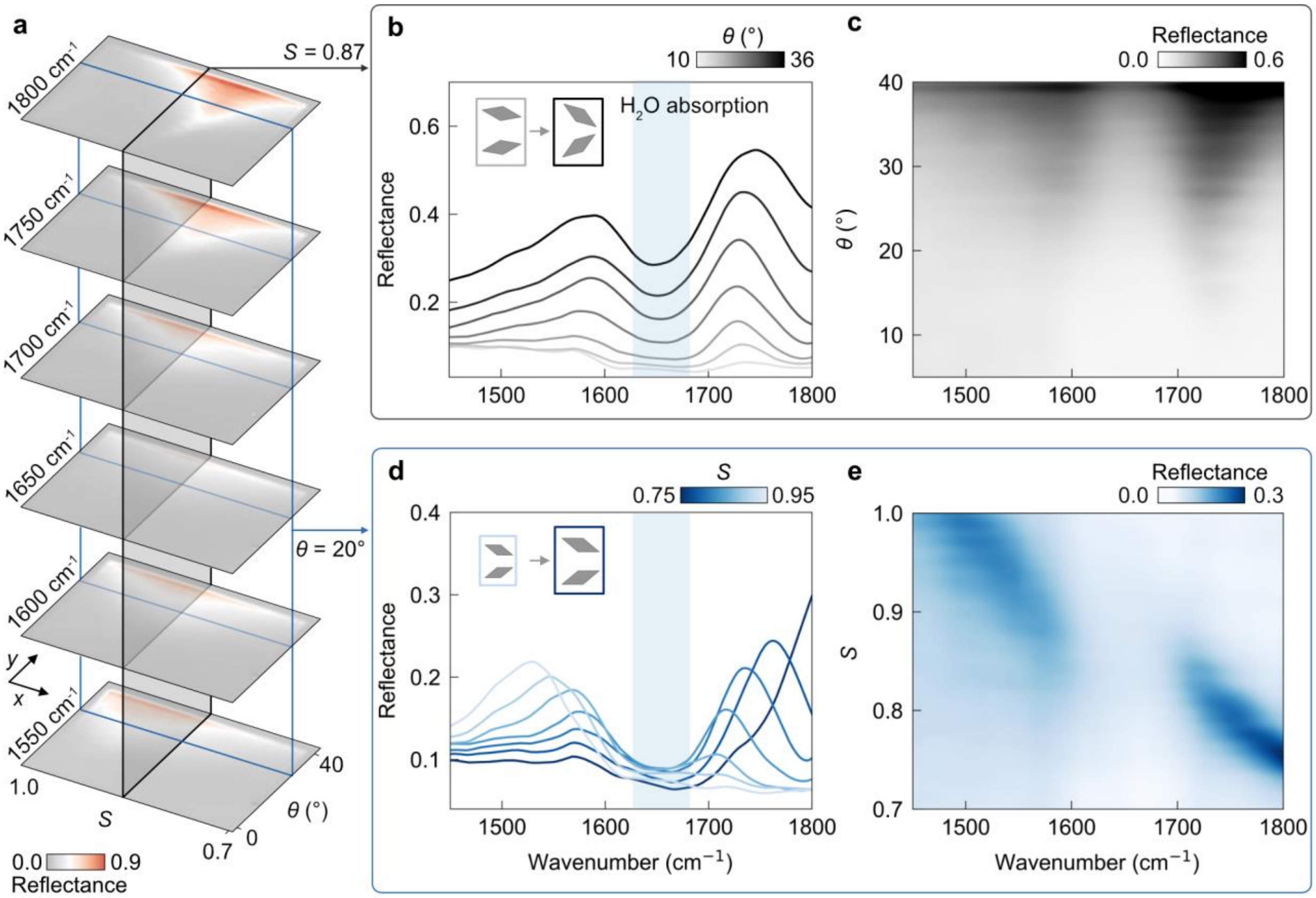

**Figure 3.** Experimental verification of a dual-gradient gold metasurface without analyte in water. (a) Reflectance snapshots measured over the wavenumber range of 1550–1800 cm$^{-1}$ with a step size of 50 cm$^{-1}$. (b) Reflectance spectra for different asymmetry parameters ($\theta$ = 10°–36°). The reflectance dip near 1650 cm$^{-1}$, highlighted by the blue shaded region, arises from water absorption. (c) Reflectance map as a function of wavenumber and asymmetry parameter, obtained by extracting data along the *y*-direction of the gradient metasurface at a fixed scaling factor *S* = 0.87. (d) Reflectance spectra for different scaling factors (*S* = 0.75–0.95). (e) Reflectance map as a function of wavenumber and scaling factor, obtained by extracting data along the x-direction of the gradient metasurface at a fixed asymmetry parameter $\theta$ = 20°.

## Real-time *In Situ* Sensing of Analyte in Water

To demonstrate the real-time *in situ* molecular sensing capability of the dual-gradient gold metasurface, we monitored the dynamics of POPC (1-palmitoyl-2-oleoyl-sn-glycero-3-phosphocholine) lipid vesicles in water. POPC was selected because it is widely used as a lipid model for cell-membrane-mimetic interfaces and molecular-interaction kinetics.[59] POPC features a characteristic carbonyl (C=O) stretching vibration at ~1735 $cm^{-1}$,[60] which serves as the target band in our sensing experiments. To promote supported lipid membrane formation, a 5 nm thick $SiO_2$ layer was deposited on the gold metasurface by electron-beam evaporation and the $SiO_2$-coated metasurface was treated with $O_2$ plasma for 30 min to render it hydrophilic, which facilitates vesicle adsorption and rupture.[61] Owing to its nanoscale thickness and negligible loss at 1735 $cm^{-1}$, the $SiO_2$ coating has only a minor influence on the qBIC optical performance (Figure S9). The metasurface was then integrated into a custom-built microfluidic system, as detailed in the Methods section, for controlled vesicle delivery and real-time monitoring of vesicle adsorption, spontaneous rupture, lipid membrane formation, and subsequent adsorption on gold resonator regions without $SiO_2$ coating.

During the molecular sensing process, the reflectance response of each gradient point across the gold metasurface was monitored in real time. A representative gradient point with $S = 0.87$ and $\theta = 20°$ exhibits a noticeable change in both reflectance amplitude and spectral position after analyte adsorption in water, arising from the dispersive optical response of POPC lipid vesicles (Figure 4a). The experimental absorbance is calculated as $A_{exp} = -\log(R_w/R_{wo})$, where $R_w$ and $R_{wo}$ denote the reflectance spectra with and without vesicles, respectively (Figure 4b). The resulting spectrum contains both the molecular vibrational response and a broadband background contribution caused by local refractive-index changes during analyte-water exchange near the gold resonators. The background component $A_b$, fitted by the asymmetric least-squares method,[22,62] is subtracted to obtain the baseline-corrected absorbance $A_{bc}$, from which the vibrational signature of the lipid vesicles is resolved (Figure 4c). An absorbance feature appears at ~1735 $cm^{-1}$ and grows over time, corresponding to the carbonyl stretching vibration of the POPC lipid vesicles (Figure 4c).

To quantify the molecular response, the integrated absorbance $\beta$ is obtained by integrating $A_{bc}$ over the lipid vibrational band between wavenumbers $\nu_1$ = 1700 $cm^{-1}$ and $\nu_2$ = 1760 $cm^{-1}$. The real-time evolution of the integrated absorbance shows a rapid initial increase followed by gradual saturation (Figure 4d). The initial rise within 10 min is attributed to vesicle adsorption, rupture, and the formation of a continuous supported lipid bilayer on the hydrophilic $SiO_2$-coated gold metasurface and substrate.[36,37] Because the sidewalls of the gold resonators are not fully coated with $SiO_2$, vesicle adsorption in these regions proceeds more gradually, leading to a slower approach to signal saturation at ~40 min. This behavior is consistent with previous quartz crystal microbalance with dissipation studies of vesicle adsorption.[63] The observed temporal evolution demonstrates direct, label-free tracking of molecular adsorption dynamics under strongly absorbing aqueous conditions.

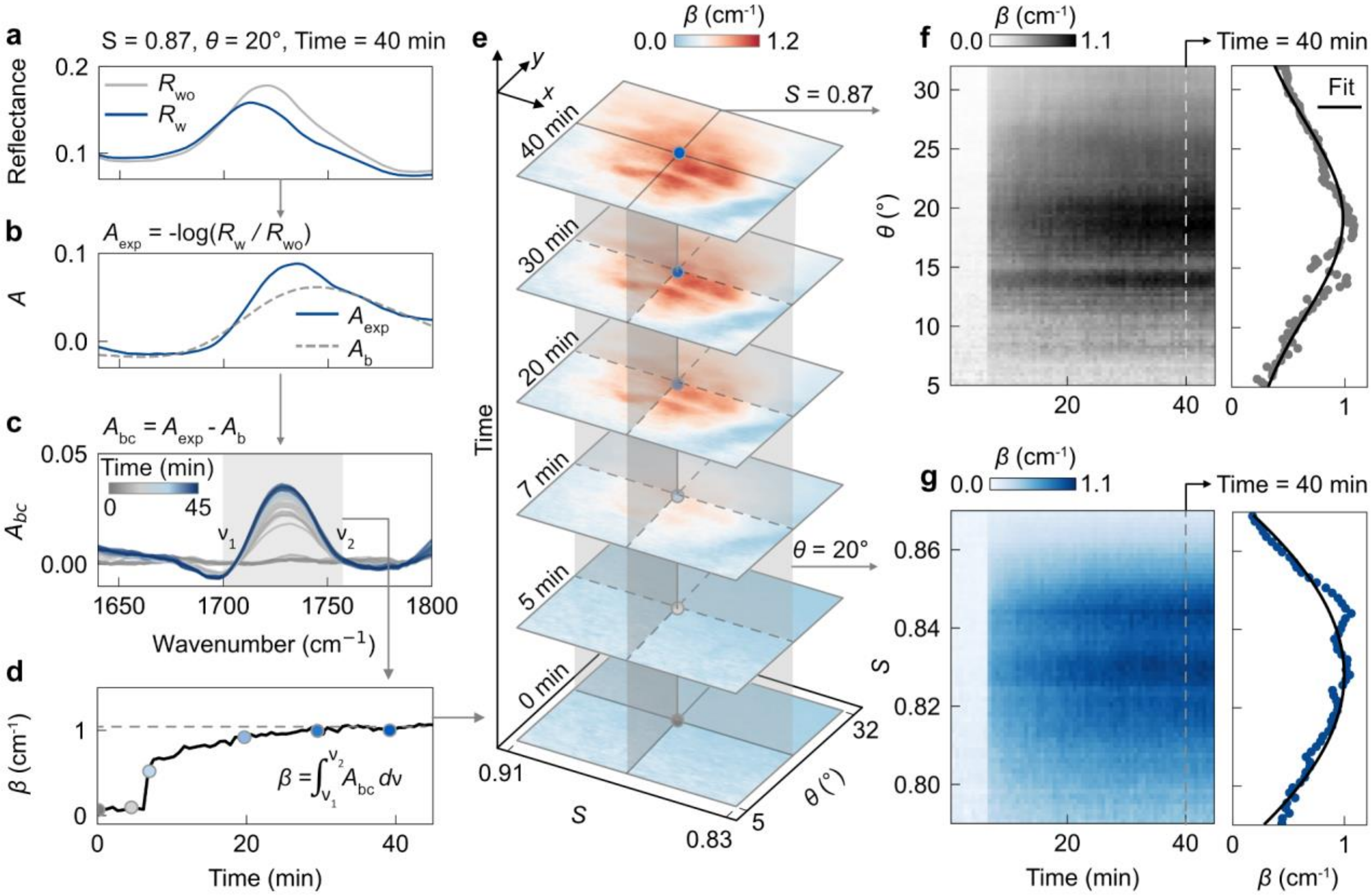

**Figure 4.** Experimental realization of real-time molecular vibrational sensing in water. (a) Reflectance spectra of a representative set of parameters ($S = 0.87$ and $\theta = 20°$) of the metasurface measured at 0 min in the absence of lipid vesicles ($R_{wo}$) and at 40 min in the presence of the lipid vesicles ($R_w$). (b) Absorbance spectrum $A_{exp}$ calculated from $R_w$ and $R_{wo}$. (c) Baseline-corrected absorbance over time, $A_{bc} = A_{exp} - A_b$, where $A_b$ is obtained using an asymmetric least-squares fitting method. A pronounced signal modulation is observed at ~1735 cm$^{-1}$, corresponding to the carbonyl vibrational band. (d) Real-time evolution of the integrated absorbance $\beta$, obtained by integrating $A_{bc}$ over the analyte vibrational band between $v_1 = 1700$ cm$^{-1}$ and $v_2 = 1760$ cm$^{-1}$ (gray shaded region in (c)). (e) Real-time evolution of the integrated absorbance $\beta$ across the dual-gradient gold metasurface. Colored markers correspond to the chosen gradient point ($S = 0.87$ and $\theta = 20°$). (f, g) Integrated absorbance $\beta$ as a function of time extracted from (e) at $S = 0.87$ (f) and $\theta = 20°$ (g). Right panels show the integrated absorbance (dots) and Gaussian fits (solid lines) at time = 40 min.

The advantage of the dual-gradient design becomes more evident when the integrated absorbance is mapped over the entire metasurface (Figure 4e and Supplementary Video 3). As lipid vesicle adsorption progresses, the integrated absorbance $\beta$ increases and eventually converges onto a single maximum-response region in the two-dimensional parameter space ($S$, $\theta$). These *in situ* measurements indicate that the molecular absorption signal is governed not only by the amount of adsorbed analyte, but also by the spectral matching and radiative loss of qBICs. The strongest sensing performance is localized in the central region of the dual-gradient metasurface, where the qBICs spectrally align with the carbonyl vibrational band while maintaining a critical coupling regime.

To gain further insight into the optimal sensing condition, time-dependent line profiles of the integrated absorbance $\beta$ were extracted along the asymmetry-parameter and scaling-factor directions by fixing $S = 0.87$ and $\theta = 20°$, respectively (Figure 4f,g). The integrated absorbance $\beta$ increases over time along both directions. Gaussian fits to both profiles at 40 min reveal that the optimal region is centered around $S = 0.87$ and $\theta = 20°$. The experimentally observed optimal asymmetry parameter is larger than the numerically predicted value of 12.5° shown in Figure 2d, which can be attributed to excessive intrinsic losses from the aqueous environment and gold resonators, as well as fabrication-induced geometric deviations.[64] These imperfections also lead to fluctuations in the integrated absorbance $\beta$ across the diamond-shaped gold resonators. Notably, this optimal sensing region remains nearly unchanged over time (Figure S10), in good agreement with the numerical results. These results demonstrate that the dual-gradient plasmonic qBIC metasurface enables real-time monitoring of molecular kinetics in aqueous environments while identifying the globally optimized sensing condition on a single device.

## CONCLUSION

We developed a single-chip dual-gradient plasmonic qBIC metasurface for enhanced *in situ* optimization molecular vibrational sensing in water. Replacing conventional elliptical nanorods with diamond-shaped resonators yields average increases of 32.2% in the $Q$-factor and 12.5% in absorbance, improving detection of weak vibrational signatures in strongly absorbing aqueous environments. Building on this optimized unit cell, we introduced a dual-gradient metasurface in which the scaling factor and asymmetry parameter are varied along two orthogonal directions. This dual-gradient design spatially encodes a series of resonator configurations within a 1000 μm × 700 μm device, covering broad ranges of resonant wavenumber and radiative decay rate without the need for multiple separately fabricated arrays. Such a strategy directly addresses a key constraint of conventional metasurface sensing, where precise spectral overlap with target molecular vibrations often requires repeated design and fabrication iterations. To describe the time-dependent adsorption process, we employed a coverage-ratio-based effective-permittivity model, in which progressive analyte adsorption is treated as analyte-water exchange within a fixed near-field volume. This model captures the evolution of sensing performance as a function of analyte coverage and asymmetry parameter, revealing the radiative coupling condition for achieving optimal absorbance.

To demonstrate the *in situ* sensing capability of the dual-gradient gold metasurface, we experimentally monitored the dynamics of POPC lipid vesicles in water. The sensing maps reveal both the kinetics of vesicle adsorption and the optimal sensing condition consistent with our numerical predictions, where the qBIC spectral position overlaps with the carbonyl stretching mode while maintaining well-matched radiative coupling strengths. Thus, the gradient gold metasurface integrates spectral and radiative-coupling matching, together with time-resolved molecular sensing on a single chip. To our knowledge, this is the first demonstration of a globally optimized qBIC sensing condition identified during *in situ* molecular vibrational sensing in water on a single chip.

Beyond the lipid vesicle model system demonstrated here, the dual-gradient metasurface provides a broad strategy for practical biosensing in complex aqueous environments, [65–67] where molecular vibrational fingerprints often consist of multiple or weak spectral features and the optimal resonance condition is unknown a priori. The densely encoded parameter space enables different vibrational bands to be matched to distinct resonator configurations, allowing each molecular signature to be probed under its optimal spectral and coupling condition. Such on-chip optimization offers an effective strategy for enhancing weak vibrational signals and multiplexed molecular analysis within a single measurement.

The generality of this approach makes it readily applicable to point-of-care diagnostics, biofluid analysis[68], live-cell metabolism studies,[69] and drug screening under physiologically relevant conditions [70], where complex molecular compositions and dynamic biochemical processes demand robust and adaptive sensing platforms. Our work establishes a spatially optimized qBIC metasurface as a versatile framework for translating mid-IR metaphotonic sensing from proof-of-concept demonstrations toward practical analytical and clinical technologies.

## METHODS

### Numerical Simulations

Numerical simulations were performed in the frequency domain using CST Studio Suite (SIMULIA). A single unit cell was modeled with periodic boundary conditions along the in-plane directions and adaptive mesh refinement to ensure numerical convergence. The lattice periods were $P_x$ = 3600 nm and $P_y$ = 2300 nm. Each gold diamond-shaped resonator had in-plane dimensions $d_1$ = 2000 nm and $d_2$ = 700 nm, with a thickness of $h_{Au}$ = 100 nm. The substrate was modeled as lossless $CaF_2$ with a refractive index of $n$ = 1.4. The wavelength-dependent optical constants of gold were taken from the experimental data reported by Olmon et al.[71] The structure was illuminated at normal incidence from the substrate side using linearly polarized light with the electric field oriented along the $y$ axis. The gold resonators were conformally coated with a 5 nm thick analyte layer. The analyte was assigned a complex refractive index of 1.45 + 0.25i.[57] To represent sensing under aqueous conditions, a homogeneous lossy superstrate was introduced above the analyte-coated metasurface. This layer had a thickness of 1800 nm and a complex refractive index of 1.34 + 0.03i, where the real part approximates the refractive index of water in the mid-infrared range and the imaginary part accounts for its absorption. The superstrate thickness was selected to provide a sufficiently large interaction volume while minimizing artificial influences from the simulation boundary. The near-field enhancement was evaluated using CST post-processing by extracting the electric-field amplitude at the mid-height plane of the gold resonators.

### Fabrication

For fabrication of the dual-gradient gold metasurface, a 400 nm thick layer of positive electron-beam resist CSAR 62 (AR-P 6200.13, Allresist) was first spin-coated onto the substrate, followed by a conductive polymer layer (ESPACER, Showa Denko K.K.) to suppress charging during electron-beam exposure. Electron-beam lithography was performed using an eLINE Plus system (Raith) operated at an acceleration voltage of 20 kV with a 20 µm aperture and a step size of 10 nm. After exposure, the resist was developed in amyl acetate, followed by a rinse in a methyl isobutyl ketone/isopropyl alcohol solution (1:9 v/v). A 5 nm chromium adhesion layer and a 100 nm gold layer were subsequently deposited by electron-beam evaporation (Bestec GmbH). The metasurface structures were then defined by a lift-off process using Remover 1165 (Micro Resist Technology GmbH). Finally, the metasurface was coated with a 5 nm thick silica layer by electron-beam evaporation to provide a hydrophilic surface for vesicle adsorption and lipid membrane formation during sensing experiments.

**Lipid Vesicle Experiments and Optical Characterization**

POPC lipid vesicles were prepared following a modified protocol reported by Nair et al.[72] Briefly, dried POPC lipid vesicles were rehydrated in 1.0 mL phosphate-buffered saline, followed by vortex mixing and bath sonication for 30 min to obtain a homogeneous vesicle suspension. Prior to the experiments, the metasurface chips were treated with oxygen plasma for 30 min to increase surface hydrophilicity and promote vesicle adsorption. For time-resolved measurements, the metasurface chip was mounted upside down in a custom-built microfluidic chamber, allowing measurements through the backside of the $CaF_2$ substrate while minimizing the optical path length in water and thereby reducing infrared attenuation. The microfluidic chamber was connected to a syringe-pump system for controlled analyte injection. The POPC vesicle suspension was introduced at a flow rate of 15 µL $min^{-1}$. Hyperspectral measurements were performed using a Spero infrared microscope (Daylight Solutions Inc., USA) equipped with a 4× objective lens (NA = 0.15), providing a field of view of approximately 2 × 2 $mm^2$. Reflectance spectra were acquired in the spectral range from 1300 to 1800 $cm^{-1}$ with a spectral resolution of 2 $cm^{-1}$. Before each experiment, a gold mirror was measured as a reference for reflectance calibration. The hyperspectral data were processed using ChemVision software for background correction, and custom Python scripts were subsequently employed to extract the metasurface spectra and perform further spectral analysis.

## ASSOCIATED CONTENT

### Supporting Information Available:

Numerical comparison of elliptical and diamond-shaped resonators, numerical background correction, spectral alignment of qBICs, experimental characterization of the gold metasurface in air and water, reflectance at different scaling factors and asymmetry parameters, effect of the $SiO_2$ layer, evolution of the integrated absorbance, and supplementary videos of the reflectance response and integrated absorbance.

## AUTHOR INFORMATION

### Corresponding Author

**Yohan Lee** - Institute of Photonics, Hamburg University of Technology, 21073 Hamburg, Germany; Chair in Hybrid Nanosystems, Nanoinstitute Munich, Faculty of Physics, Ludwig-Maximilians-University, 80539 Munich, Germany; Email: yohan.lee@tuhh.de

**Alexander A. Antonov** - Institute of Photonics, Hamburg University of Technology, 21073 Hamburg, Germany; Chair in Hybrid Nanosystems, Nanoinstitute Munich, Faculty of Physics, Ludwig-Maximilians-University, 80539 Munich, Germany; Email: alexander.antonov@tuhh.de

**Andreas Tittl** - Institute of Photonics, Hamburg University of Technology, 21073 Hamburg, Germany; Chair in Hybrid Nanosystems, Nanoinstitute Munich, Faculty of Physics, Ludwig-Maximilians-University, 80539 Munich, Germany; Email: andreas.tittl@tuhh.de

### Authors

**Tao Jiang** - Institute of Photonics, Hamburg University of Technology, 21073 Hamburg, Germany; Chair in Hybrid Nanosystems, Nanoinstitute Munich, Faculty of Physics, Ludwig-Maximilians-University, 80539 Munich, Germany

**Michael Hirler** - Institute of Photonics, Hamburg University of Technology, 21073 Hamburg, Germany; Chair in Hybrid Nanosystems, Nanoinstitute Munich, Faculty of Physics, Ludwig-Maximilians-University, 80539 Munich, Germany

**Lina Rohrer** - Institute of Photonics, Hamburg University of Technology, 21073 Hamburg, Germany; Chair in Hybrid Nanosystems, Nanoinstitute Munich, Faculty of Physics, Ludwig-Maximilians-University, 80539 Munich, Germany

**Martin Barkey** - Chair in Hybrid Nanosystems, Nanoinstitute Munich, Faculty of Physics, Ludwig-Maximilians-University, 80539 Munich, Germany

**Dmytro Gryb** - Chair in Hybrid Nanosystems, Nanoinstitute Munich, Faculty of Physics, Ludwig-Maximilians-University, 80539 Munich, Germany

**Leonardo de S. Menezes** - Chair in Hybrid Nanosystems, Nanoinstitute Munich, Faculty of Physics, Ludwig-Maximilians-University, 80539 Munich, Germany; Departamento de Física, Universidade Federal de Pernambuco, 50670-901 Recife-PE, Brazil

**Silvia Holler** - Cellular Computational and Integrative Biology Department, University of Trento, 38123 Trento, Italy

**Stefan A. Maier** - School of Physics and Astronomy, Monash University, Clayton, 3800 Victoria, Australia; Department of Physics, Imperial College London, SW7 2BW London, UK

**Author Contributions**

T.J., Y.L., A.A.A., and A.T. contributed to the conceptualization, data collection, and writing of the manuscript, developed the methodology, and performed the data analysis. S.H. prepared and provided the lipid vesicle samples. M.H., L.R., D.G., L.S.M., S.H., S.A.M., Y.L., A.A.A., and A.T. contributed to the review and editing of the manuscript. S.A.M., Y.L., A.A.A., and A.T. supervised the project. All authors reviewed and approved the final manuscript and agree to be accountable for all aspects of the work, ensuring its accuracy and integrity.

**Notes**

The authors declare no competing financial interest.

## ACKNOWLEDGMENT

This project was funded by the Deutsche Forschungsgemeinschaft (DFG, German Research Foundation) under Germany's Excellence Strategy EXC 3120/1 – 533771286 and EXC 2089/1–390776260 and the Emmy Noether Programme (TI 1063/1), the Bavarian program Solar Energies Go Hybrid (SolTech), and Enabling Quantum Communication and Imaging Applications (EQAP), and the Center for NanoScience (CeNS). It was also funded by the European Union (ERC, METANEXT, 101078018 and EIC, OMICSENS, 101129734). The views and opinions expressed are, however, those of the authors only and do not necessarily reflect those of the European Union, the European Research Council Executive Agency, or the European Innovation Council and SMEs Executive Agency (EISMEA). Neither the European Union nor the granting authority can be held responsible for them. S.A.M. additionally acknowledges the Lee-Lucas Chair in Physics.

## Supporting Information for

# Dual-Gradient Plasmonic qBIC Metasurface for Time-Resolved *In Situ* Optimization of Molecular Vibrational Sensing in Water

Tao Jiang[1,2], Michael Hirler[1,2], Lina Rohrer[1,2], Martin Barkey[2], Dmytro Gryb[2], Leonardo de S. Menezes[2,3], Silvia Holler[4], Stefan A. Maier[5,6], Yohan Lee*[1,2], Alexander A. Antonov†[1,2], Andreas Tittl‡[1,2]

[1]Institute of Photonics, Hamburg University of Technology, 21073 Hamburg, Germany

[2]Chair in Hybrid Nanosystems, Nanoinstitute Munich, Faculty of Physics, Ludwig-Maximilians-University, 80539 München, Germany

[3]Departamento de Física, Universidade Federal de Pernambuco, 50670-901 Recife-PE, Brazil

[4]Cellular Computational and Integrative Biology Department,

University of Trento, 38123 Trento, Italy

[5]School of Physics and Astronomy, Monash University, Clayton, 3800 Victoria, Australia

[6]Department of Physics, Imperial College London, SW7 2BW London, UK

Email: * yohan.lee@tuhh.de / † alexander.antonov@tuhh.de / ‡ andreas.tittl@tuhh.de

## Contents

**Supplementary Note 1: Numerical Comparison of Elliptical and Diamond-Shaped Resonators**

To numerically compare the sensing performance of elliptical and diamond-shaped resonators, we align their qBIC positions at 1730 $cm^{-1}$ for asymmetry parameters ranging from 7.5° to 35°. Figure S1a shows the corresponding reflectance spectra without and with analyte at $\theta$ = 15°, 25°, and 35°. The analyte was modeled as a 5 nm thick conformal layer covering the metasurface. The maximum reflectance extracted from each spectrum is shown in Figure S1b. As the asymmetry parameter increases, the reflectance amplitude increases for both resonator geometries, whereas the absorbance peaks at $\theta$ = 12.5° (Figure S1c). Across this range, the diamond-shaped resonators exhibit higher absorbance than the elliptical resonators, with an 8.9% relative increase at $\theta$ = 12.5° and an average relative increase of 12.5% over the asymmetry range from 7.5° to 35° (Figure S1d). The relative absorbance increase is calculated as $(A_{\text{diamond}} - A_{\text{ellipse}})/A_{\text{ellipse}} \times 100\%$, where $A_{\text{diamond}}$ and $A_{\text{ellipse}}$ are the absorbances of the diamond-shaped and elliptical resonators, respectively. A stronger enhancement is observed for the $Q$-factor, which increases by 30.8% from 11.7 to 15.3 at $\theta$ = 12.5° and by 32.2% on average over the same asymmetry range (Figure S1e,f). The relative $Q$-factor increase is given by $(Q_{\text{diamond}} - Q_{\text{ellipse}})/Q_{\text{ellipse}} \times 100\%$, where $Q_{\text{diamond}}$ and $Q_{\text{ellipse}}$ are the $Q$-factors of the diamond-shaped and elliptical resonators, respectively.

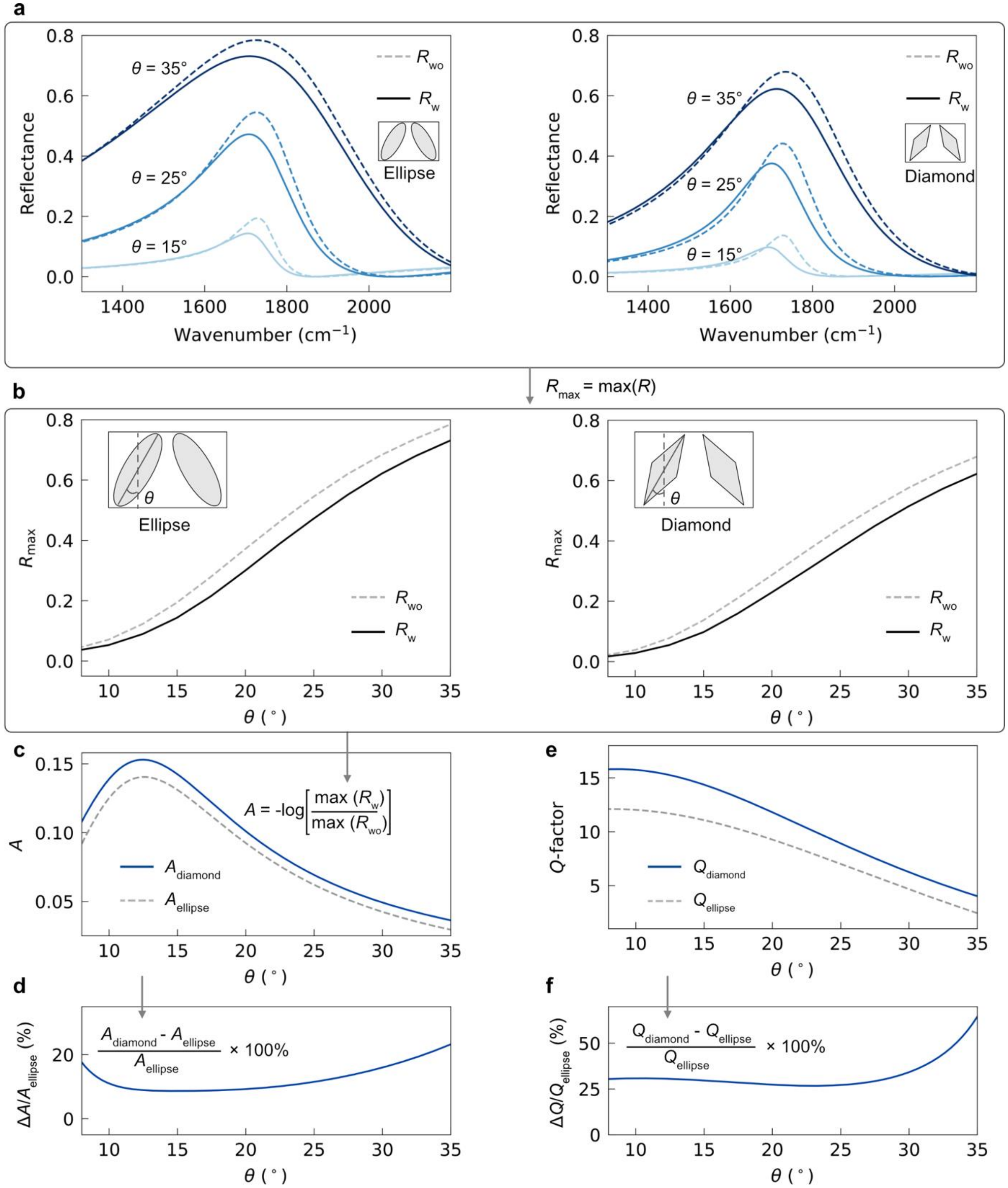

**Figure S1.** Comparison of elliptical and diamond-shaped resonators. (a) Reflectance spectra of the metasurfaces without and with analyte. (b) Maximum reflectance as a function of the asymmetry parameter $\theta$ for both conditions. (c) Absorbance and (d) corresponding relative increase for the elliptical and diamond-shaped resonators. (e) $Q$-factor and (f) corresponding relative increase for both geometries.

**Supplementary Note 2: Numerical Analysis of Background Correction**

During the sensing process, the analyte covers the gold metasurface and forms a conformal thin layer on the gold nanorods. Figure S2a schematically illustrates the metasurface in water before and after analyte adsorption. The analyte permittivity was modeled using an adapted Lorentzian dispersion model.[1] Once the analyte layer is introduced, the reflectance spectrum changes not only in amplitude, but also in the spectral position of the qBIC (Figure S2b). This spectral shift arises from the difference between the dispersive optical properties of the analyte and those of water, which modifies the local refractive-index environment of the resonators. Figure S2c shows that the calculated absorbance contains both a narrow molecular vibrational contribution indicated by the gray shaded region and a broadband background. The combined absorbance exhibits a step-like spectral profile,[2] which obscures the molecular vibrational band and complicates direct quantification of the analyte absorption.

To remove this background, we exploit the asymmetric least-squares (ALS) method.[3,4] The slowly varying spectral background $A_\mathrm{b}$ is obtained by minimizing a weighted least-squares function combined with a second-difference smoothness penalty:[5,6]

$$\sum_{i=1}^{N} w_\mathrm{i}\left(A_{\mathrm{sim},i} - A_{\mathrm{b},i}\right)^2 + m\sum_{i=1}^{N-2}\left(\Delta^2 A_{\mathrm{b},i}\right)^2 \tag{S1}$$

Here, $A_{\mathrm{sim},i}$ is the simulated absorbance at the $i$-th wavenumber, $A_{\mathrm{b},i}$ is the fitted background at the same wavenumber, and $\Delta^2 A_{b,i} = A_{b,i+1} - 2A_{b,i} + A_{b,i-1}$ is the second finite difference of the fitted background, which quantifies the local curvature. $N$ is the total number of spectral data points used in the ALS fitting, $w_\mathrm{i}$ is the weighting factor assigned to the $i$-th data point, and $m$ is the smoothing parameter.

The first term describes the weighted deviation between the simulated absorbance and the fitted background, while the second term constrains the curvature of the fitted background. The smoothing parameter $m$ determines the balance between spectral fidelity and background smoothness. A larger $m$ imposes stronger smoothing and produces a flatter background, whereas a smaller $m$ allows the fitted background to follow local spectral variations more closely and may lead to overfitting of the molecular vibrational feature.

During each iteration, data points above the fitted background are assigned a weight of $p$, whereas those below the background are assigned a weight of $1 - p$:

$$w_i = \begin{cases} p, & A_{sim,i} > A_{b,i} \\ 1 - p, & A_{sim,i} \leq A_{b,i} \end{cases} \tag{S2}$$

A smaller value of $p$ assigns less weight to data points above the fitted background and therefore keeps the baseline below the positive absorption feature. The parameters $m = 5 \times 10^3$ and $p = 0.05$ were selected, which yield a good fit to the background signal.

After background subtraction, the analyte absorbance exhibits a distinct vibrational peak at approximately 1730 $\mathrm{cm}^{-1}$ (Figure S2d). Two smaller side peaks are also observed on either side of the main band. These features are attributed to local overfitting of the background in the adjacent spectral regions rather than to additional molecular vibrations.

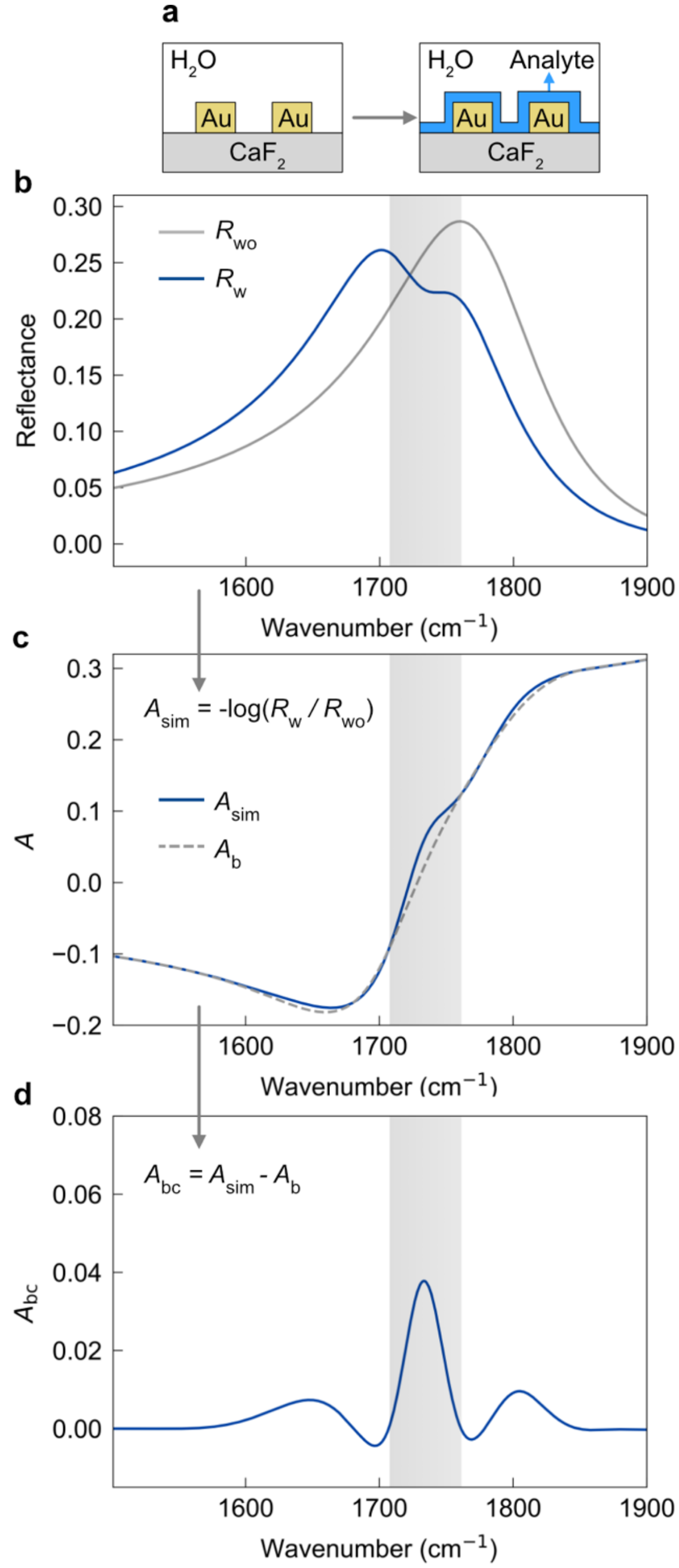

**Figure S2.** Numerical analysis of background correction. (a) Schematic of the metasurface in water with and without analyte. (b) Simulated reflectance spectra of the metasurface without and with analyte. (c) Absorbance $A_{sim}$ and the fitted background component $A_b$ obtained using the asymmetric least-squares method. (d) Baseline-corrected absorbance $A_{bc}$. The gray shaded region indicates the analyte vibrational band.

**Supplementary Note 3: Spectral Alignment of qBICs**

Increasing the asymmetry parameter $\theta$ enhances the radiative coupling of the qBIC, leading to a broader linewidth and higher amplitude. It also shifts its spectral position by changing the coupling between the nanorods and the corresponding resonance condition (Figure S3a). Thus, $\theta$ influences both the radiative coupling strength and the spectral position of the qBIC.

To decouple these two effects, we spectrally align the qBICs by introducing a local alignment factor $S_a$, which is applied to both the long axis $d_1$ and the short axis $d_2$ of the diamond-shaped nanorods (Figure S3b,c). We first determine $S_a$ numerically at several discrete values of $\theta$ and then fit the resulting data using a third-order polynomial to obtain $S_a$ at finer $\theta$ intervals. The fitted curve shows excellent agreement with the simulated data (Figure S3c). The simulated reflectance spectra in water are consequently aligned at the same spectral position. This procedure preserves the asymmetry parameter $\theta$ while compensating for the spectral shift through the alignment factor $S_a$, allowing the effect of $\theta$ on the radiative coupling strength to be evaluated independently of the resonance shift.

After spectral alignment, the reflectance amplitudes are higher than those before alignment. This enhancement is advantageous for biosensing in water, since the stronger radiative coupling maintains the far-field amplitude of the qBIC against the substantial nonradiative damping introduced by water absorption. The resonance remains more readily detectable in the highly lossy aqueous environment, enabling real-time tracking of spectral changes caused by analyte adsorption.

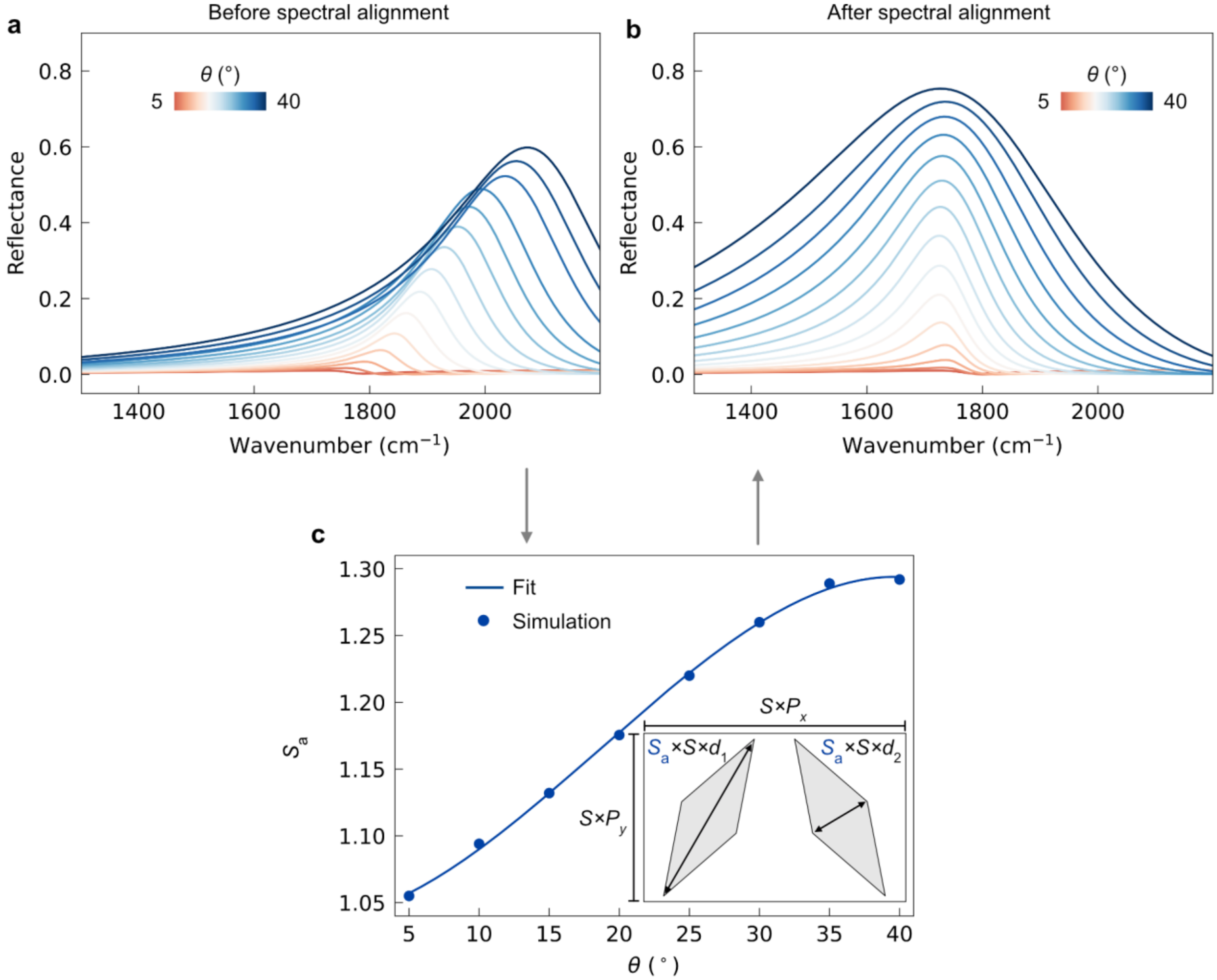

**Figure S3.** Alignment of qBIC spectral positions. Simulated reflectance spectra (a) before and (b) after spectral alignment of the qBICs. (c) Alignment curve for the diamond-shaped resonators. A local alignment factor $S_a$ is applied to the lateral dimensions $d_1$ and $d_2$. $S$ is the scaling factor and $P_x$ and $P_y$ are the lattice periods, as defined in the main text. The alignment curve is fitted as $-6.98 \times 10^{-6}\,\theta^3 + 3.898 \times 10^{-4}\,\theta^2 + 1.964 \times 10^{-3}\,\theta + 1.038$.

**Supplementary Note 4: Experimental Performance of the Gold Metasurface in Air**

The metasurface was characterized in air using a Spero microscope equipped with a quantum cascade laser (QCL), which records a reflectance spectrum at each spatial position (Figure S4). At a fixed wavenumber selected from the QCL sweep, the resonant response in air appears at larger scaling factors than in water (Figure 3). This shift arises from the lower refractive index of air, which blueshifts the qBIC spectral position. Therefore, resonators with larger scaling factors in air are required to restore the resonance condition at the same excitation wavenumber.

This response also demonstrates the potential of the dual-gradient gold metasurface for refractive-index sensing.[7–9] The underlying principle is that changes in the surrounding refractive index alter the optical environment of the resonators and shift the qBIC spectral position. Through the scaling-factor gradient, continuous qBIC spectral information is spatially encoded across the metasurface, which converts spectral shifts induced by refractive index changes into spatial displacements of the resonant response.[10] This enables direct spatial readout and facilitates compact, label-free detection and real-time monitoring of changes in the local dielectric environment.

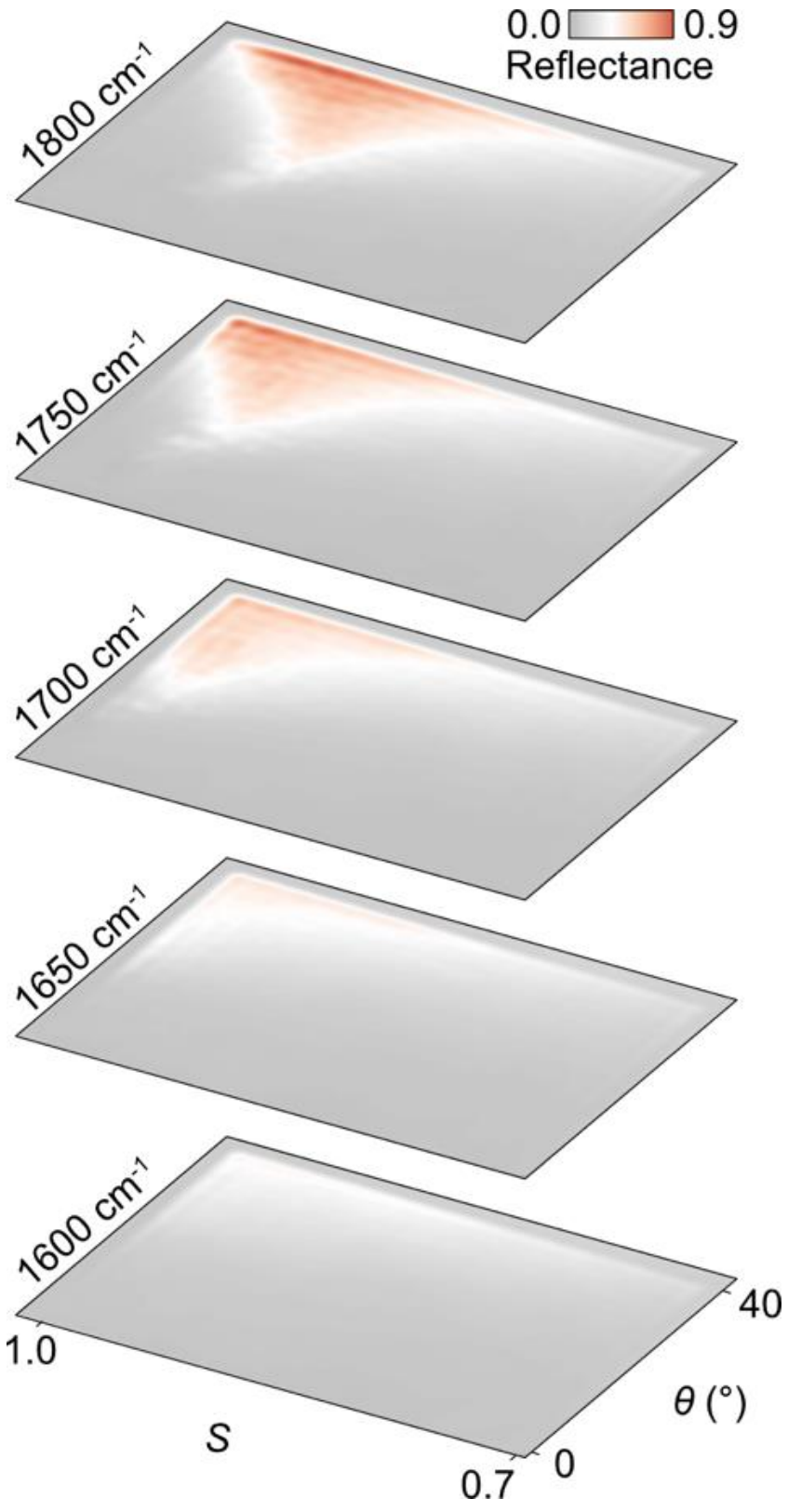

**Figure S4.** Reflectance snapshots measured in air over the wavenumber range of 1600–1800 $cm^{-1}$ with a step size of 50 $cm^{-1}$.

## Supplementary Note 5: Reflectance in Air at Different Scaling Factors

Experimental reflectance spectra of the metasurface in air were extracted at scaling factors of S = 0.95, 0.97, and 0.99 (Figure S5). Unlike the performance in water, the qBIC spectral positions are no longer aligned across different asymmetry parameteres. Similarly, the reflectance maps (Figure S5d–f) exhibit a tilted resonance band, rather than the nearly vertical resonance band observed in water.

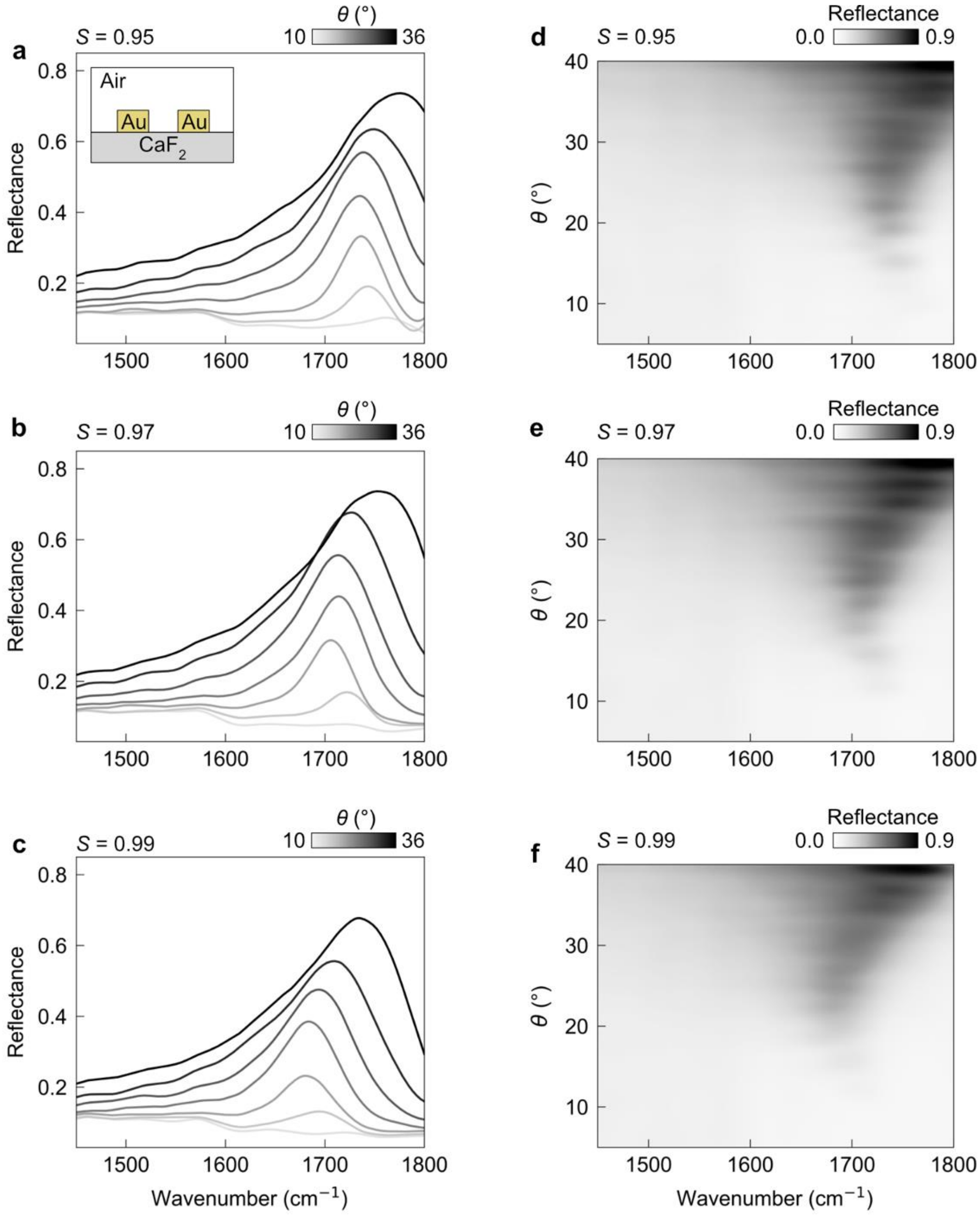

**Figure S5.** Experimental reflectance spectra in air at fixed scaling factors of (a) $S = 0.95$, (b) $S = 0.97$, (c) $S = 0.99$, with the corresponding reflectance maps as functions of wavenumber and asymmetry parameter shown in (d–f), respectively. The qBICs are not spectrally aligned,

particularly at large asymmetry parameters, since the spectral alignment is designed for the water environment.

### Supplementary Note 6: Reflectance in Water at Different Scaling Factors

Experimental reflectance spectra of the metasurface measured in water were extracted at scaling factors of $S$ = 0.84 and 0.90 (Figure S6). The qBIC remains clearly visible for most asymmetry parameters, except near $\theta$ = 10°, where the reflectance is weak due to low radiative coupling. Interaction with the water absorption band near 1650 cm$^{-1}$ gives rise to two spectral features on either side of the absorption region. Increasing the scaling factor from $S$ = 0.84 to 0.90 redshifts the qBIC and brings it into closer spectral overlap with the water absorption.

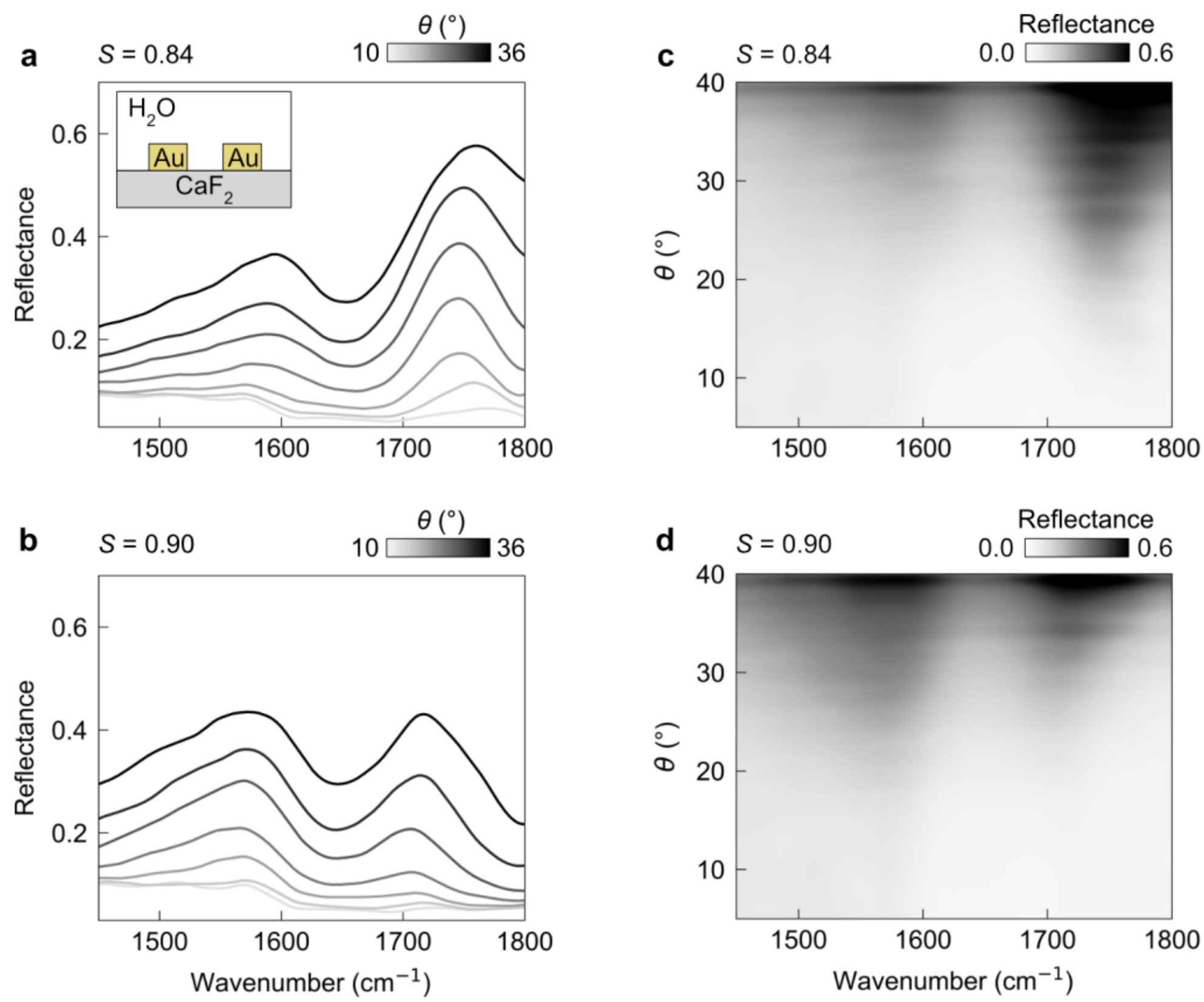

**Figure S6.** Experimental reflectance spectra in water at fixed scaling factors of (a) S = 0.84, (b) S = 0.90, with the corresponding reflectance maps as functions of wavenumber and asymmetry parameter shown in (c,d), respectively.

### Supplementary Note 7: Reflectance in Air at Different Asymmetry Parameters

Experimental reflectance spectra of the metasurface measured in air were extracted at asymmetry parameters of $\theta$ = 10°, 20°, 30° and 40° (Figure S7). The reflectance increases from below 0.15 at $\theta$ = 10° to nearly unity at $\theta$ = 40°, accompanied by a substantial reduction in the $Q$-factor.

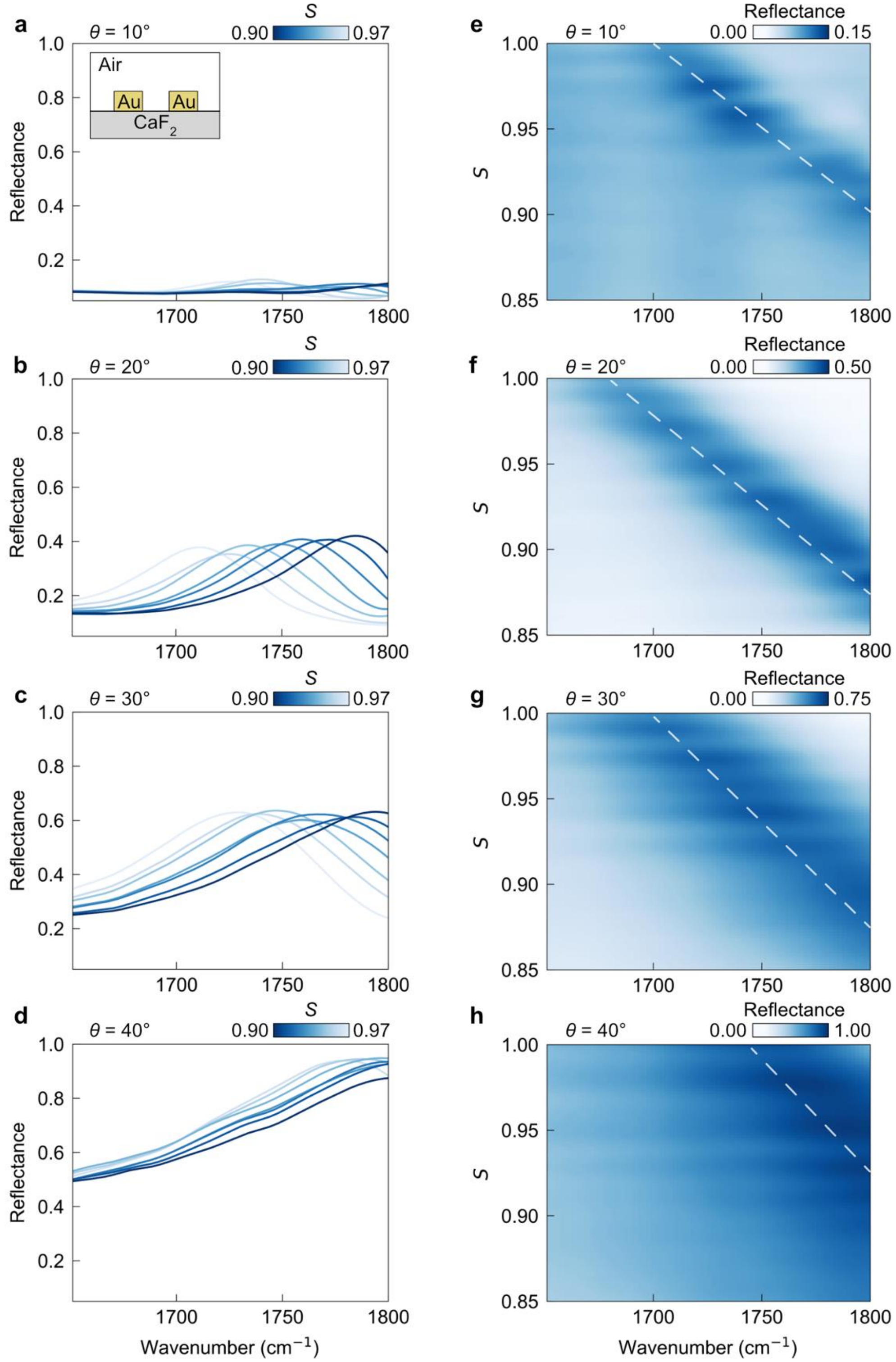

**Figure S7.** Experimental reflectance spectra in air at fixed asymmetry parameters of (a) $\theta = 10°$, (b) $\theta = 20°$, (c) $\theta = 30°$, (d) $\theta = 40°$, with the corresponding reflectance maps as functions of wavenumber and scaling factor shown in (e–h), respectively. White lines indicate the qBIC positions.

## Supplementary Note 8: Reflectance in Water at Different Asymmetry Parameters

Experimental reflectance spectra of the metasurface in water were extracted at asymmetry parameters of $\theta$ = 10°, 30°, and 40° (Figure S8). With increasing $\theta$, stronger radiative coupling raises the overall reflectance, thereby reducing the relative contrast of the water-absorption dip near 1650 cm$^{-1}$ and making it less distinct (Figure S8d–f).

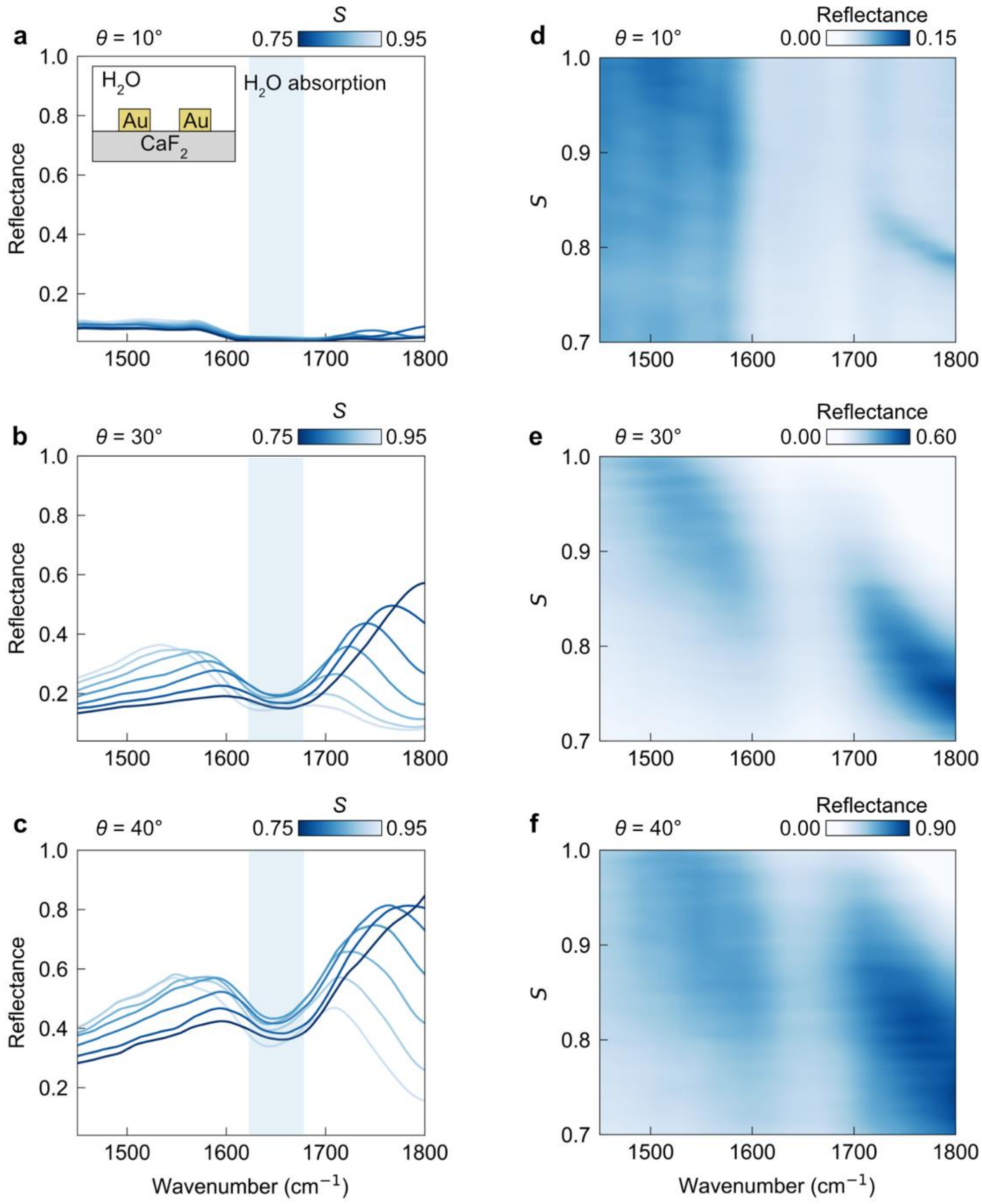

**Figure S8.** Experimental reflectance spectra in water at fixed asymmetry parameters of (a) $\theta$ = 10°, (b) $\theta$ = 30°, (c) $\theta$ = 40°, with the corresponding reflectance maps as functions of wavenumber and scaling factor shown in (d–f).

## Supplementary Note 9: Effect of the $SiO_2$ Layer

To evaluate the influence of the $SiO_2$ coating deposited by electron-beam evaporation, we simulated the metasurface with a 5 nm thick $SiO_2$ layer deposited on the top surfaces of the gold nanorods and on the exposed $CaF_2$ substrate (Figure S9). The reflectance spectra with and without the $SiO_2$ layer are nearly identical. The minor variation in reflectance amplitude is attributed to the weak dielectric perturbation introduced by the ultrathin $SiO_2$ coating and to numerical deviations associated with mesh discretization.

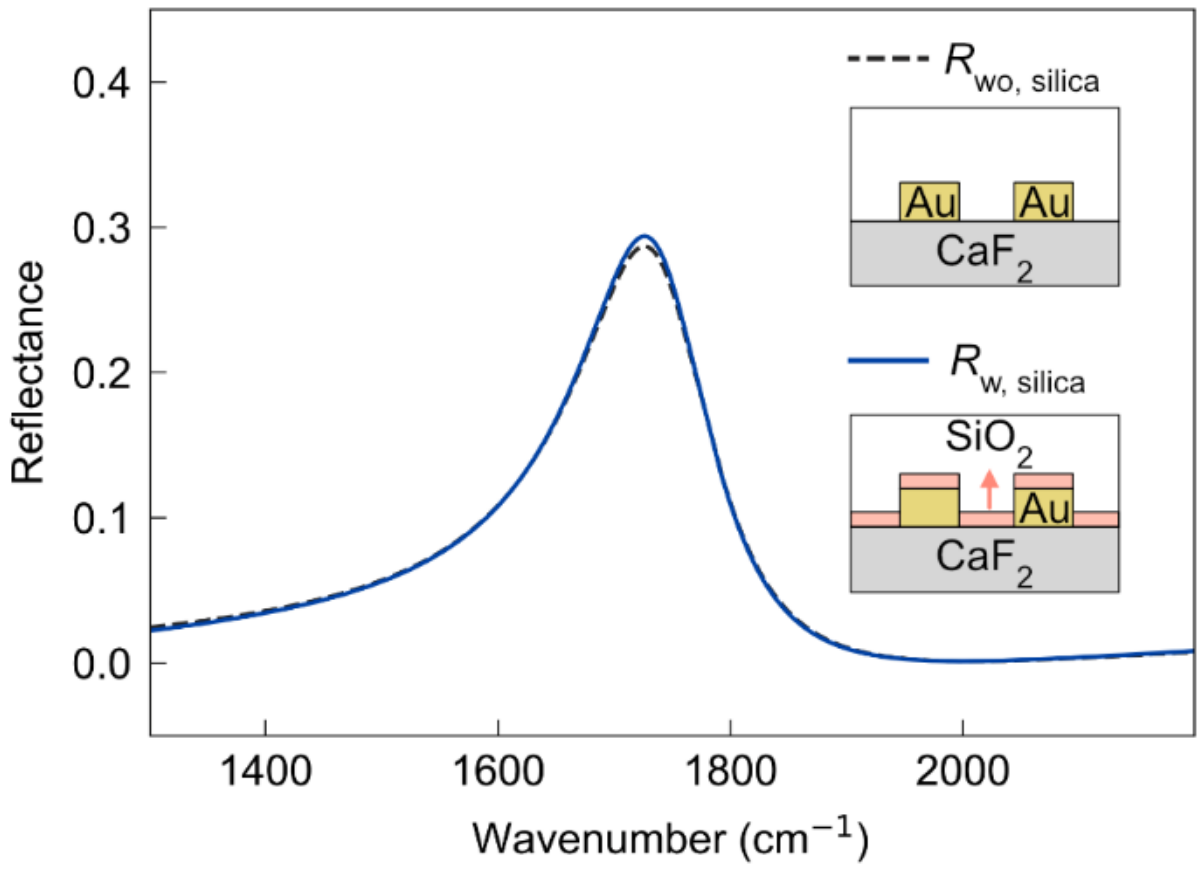

**Figure S9.** Simulated reflectance spectra of the metasurface without and with a $SiO_2$ layer of 5 nm in thickness. The insets show schematics of the gold metasurface without and with the $SiO_2$ layer, highlighted in pink.

## Supplementary Note 10: Evolution of Integrated Absorbance

Two gradient line cuts, along the scaling-factor gradient at $\theta = 20°$ and the asymmetry-parameter gradient at $S = 0.87$, are selected to illustrate the evolution of $\beta$ from 0 to 30 min at 10 min intervals (Figure S10). After 10 min, $\beta$ exhibits an approximately Gaussian distribution along both gradient directions. The two profiles reach their respective maxima at approximately $S = 0.87$ and $\theta = 20°$. These peak positions remain nearly unchanged throughout the measurement, while the integrated absorbance continuously increases until 30 min, consistent with the numerical predictions presented in the main text (Figure 2).

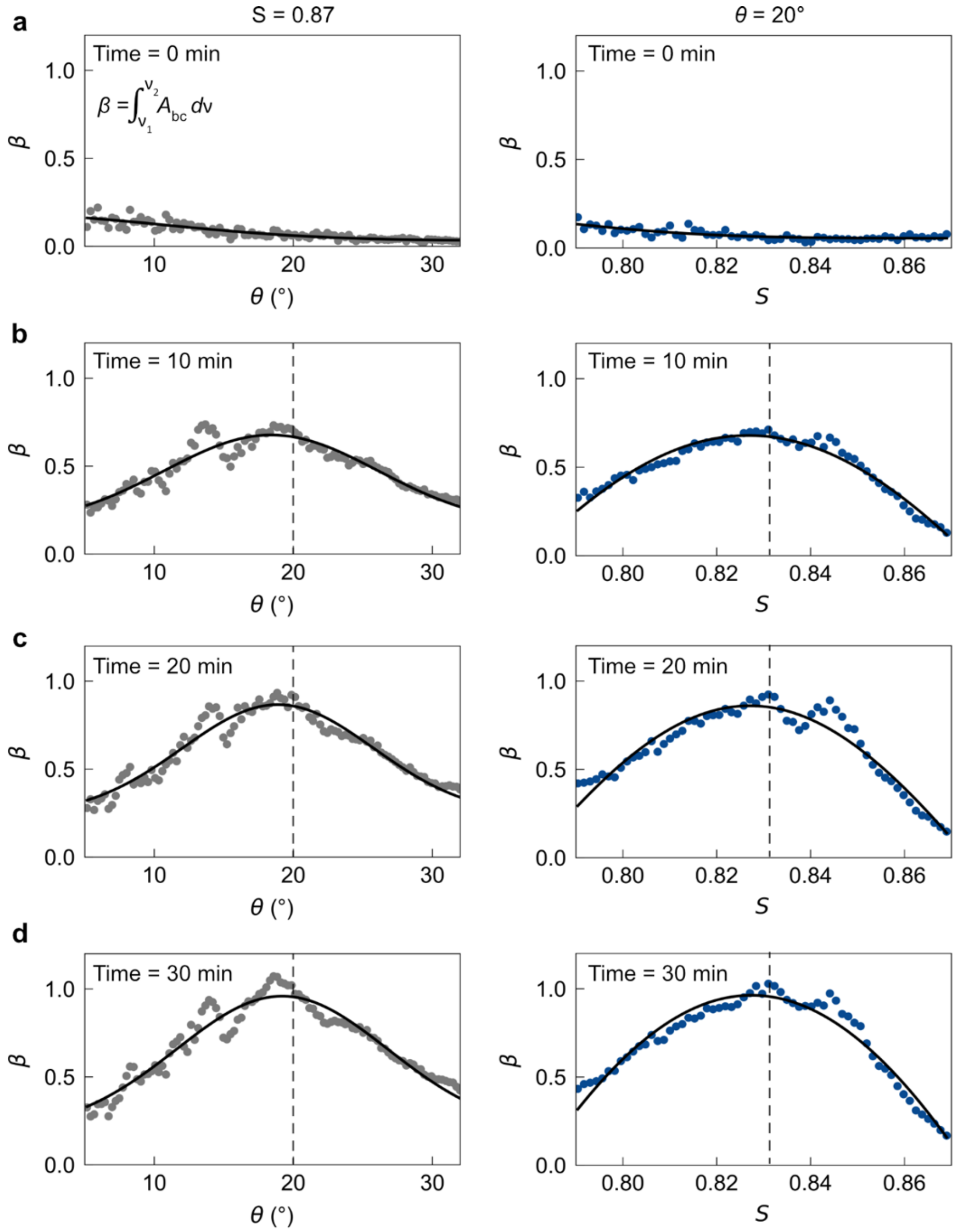

**Figure S10**. Experimental integrated absorbance $\beta$ (dots) and corresponding Gaussian fits (solid lines) at (a) time = 0 min, (b) time = 10 min, (c) time = 20 min, and (d) time = 30 min. The left and right panels show $\beta$ at a fixed scaling factor of $S = 0.87$ and asymmetry parameter of $\theta = 20°$, respectively.

**Supplementary Video 1: Reflectance video of the Metasurface in Air**

The fabricated metasurface spans scaling factors $S = 0.7–1.0$ and asymmetry parameters $\theta = 0°–40°$, and is characterized using a hyperspectral imaging microscope. Each frame corresponds to a single wavenumber, increasing from 1500 to 1800 $cm^{-1}$.

**Supplementary Video 2: Reflectance video of the Metasurface in Water**

Reflectance of the dual-gradient gold metasurface measured in water, corresponding to Figure 3 in the main text. The metasurface is integrated into a custom-built microfluidic cell, through which water is introduced, while the reflectance is collected from the substrate side. Frames are displayed sequentially from 1500 to 1800 $cm^{-1}$, with each frame representing a single wavenumber.

**Supplementary Video 3: Real-time Evolution of Integrated Absorbance**

Real-time evolution of the integrated absorbance across the dual-gradient gold metasurface, corresponding to Figure 4 in the main text. The analyzed region is defined by scaling factors $S = 0.79–0.87$ and asymmetry parameters $\theta = 5°–32°$. The sequence of frames shows the spatial distribution of the integrated absorbance throughout an approximately 40 min measurement period.